\documentclass{llncs}

\usepackage{graphicx}
\usepackage{subfigure}

\newcommand{\iseij}[6]{\left\{
    \begin{array}{ccc}
      {#1}&{#2}&{#3}\\
      {#4}&{#5}&{#6}
    \end{array}\right\}}

\title{Symmetric angular momentum coupling, the quantum volume operator and the 7-spin network: a computational perspective}
\titlerunning{7-spin networks}
\author{Dimitri Marinelli\inst{1} \and Annalisa Marzuoli\inst{2,3} \and Vincenzo Aquilanti\inst{4,5}\and
        Roger W. Anderson\inst{6} \and Ana Carla P. Bitencourt\inst{7} \and Mirco Ragni\inst{7}}
\authorrunning{Mirco Ragni}

 \institute{Dipartimento di Fisica, Universit\`a degli Studi di Pavia, Italy \and
           INFN, Sezione di Pavia, Italy \and
           Dipartimento di Matematica ``F Casorati'', Universit\`a degli Studi di Pavia, Italy \and
           Dipartimento di Chimica, Biologia e Biotecnologie,  Universit\`a di Perugia, Italy \and
           Instituto de F\'isica, Universidade Federal da Bahia, Brasil \and
           Department of Chemistry, University of California, Santa Cruz, CA 95064, U.S.A. \and
           Departamento de F\'isica, Universidade Estadual de Feira de Santana, Brazil
          }

\begin{document}

\maketitle

\begin{abstract}
A unified vision of the symmetric coupling of angular momenta and
of the quantum mechanical volume operator is illustrated. The
focus is on the quantum mechanical angular momentum theory of
Wigner's 6j symbols and on the volume operator of  the symmetric
coupling in spin network approaches: here, crucial to our
presentation are an appreciation of the role of the Racah sum rule
and the simplification arising from the use of Regge symmetry. The
projective geometry approach permits the introduction of a
symmetric representation of a network of seven spins or angular
momenta. Results of extensive computational investigations are
summarized, presented and briefly discussed.
\end{abstract}

\section{Introduction}\label{s1}

This paper extends the theory presented in
\cite{aquilanti2013volume,aquilanti2014} and provides numerical
documentation.

The attention towards the symmetric coupling of angular momenta
started  half a century ago
\cite{cha1064,jeanmarclevyleblond1965symmetrical}: in the late
Nineties, an essentially equivalent problem attracted the keen
interest in spin network theories, of great relevance in current
approaches to quantum gravity, see references in
\cite{aquilanti2013volume,aquilanti2014}, where of extreme
importance is the formulation of a quantum mechanical volume
operator and the study of its properties. The intimate connection
between the two formulations was later established \cite{car2002}.

Mathematical advances regarded relationships with continuous and
discrete orthogonal polynomials, their Askey scheme
classification, the quadratic and cubic algebras (see
\cite{granovskii19921,gen2014}), the Leonard pairs and triples
(see \cite{ter2001} and later papers by Terwilliger). Progress in
angular momentum theory has mainly regarded semiclassical
asymptotics of $3nj$ symbols
\cite{ponzregge,neville2006volume1,neville2006volume2,neville1971technique,nikiforovsuslovuvarov199110,schgora,schgorb,rbfaal.10,littlejohn2009uniform,aquilanti2007semiclassical},
also illuminating regarding the geometrical aspects. For the
latter in this paper we will particularly  make reference to
projective viewpoints
\cite{fanoracah1959,robinson1970group,judd1983angularmomentum,bilo9,Lab1975,Lab1998,Lab2000}).

Applications to atomic, molecular and nuclear physics have been
intensive; computational aspects have been discussed recently (see
previous papers in this series
\cite{bitencourt2012exact,lncs1.2013,lncs2.2013}). At times, the
investigators in these areas carried out their research without
exchanges among them and using different formalisms, making it
difficult the appreciation and utilization from communities of
applied physical, chemical and in general computational
scientists. Here we try to partially fill this gap by giving the
basic formulations, and then illustrations of the perspective
emerging from very recent progress through presentation and
discussion of computational results. Next section \ref{s2}
outlines the relevant formulation; section \ref{s3} reports
results of calculations of caustics and ridges, of interest for
semiclassical analysis, and of the potentials $U^+$ and $U^-$ for
the volume operator, which acts democratically on the 7-spin
network. Concluding remarks are in section \ref{s4} .

\section{Basic theory}\label{s2}
The focus in this section is on Racah sum rule formula and an
account follows, inspired by the hexagonal representation in Fig.
2i in Ref. \cite{ac.01}, see the related Eq. (6) there and also for
example the corresponding equations in the comprehensive
compilation \cite{varsh}. In these formulas three $6j$ symbols
appear involving four angular momenta (or spins) $j_a$, $j_b$,
$j_c$, $j_d$, also denoted when convenient simply $a$, $b$, $c$, $d$; to
them the previous convention \cite{lncs1.2013,lncs2.2013} is
conveniently applied and extended: first, the choice of proper
Regge quadrilaterals runs as before, as that of taking the set
containing the minimum of eight combinations, individuating $a$;
then, we can arrange the four spins according to $a \leq b \leq c \leq d$. We introduce here
the notation $x$, $y$, $z$ for the three intermediates,  which are the
diagonals of possible quadrilaterals.

\begin{figure}
  \center
  \includegraphics[width=\textwidth, clip, trim= .7cm .7cm .7cm .7cm]{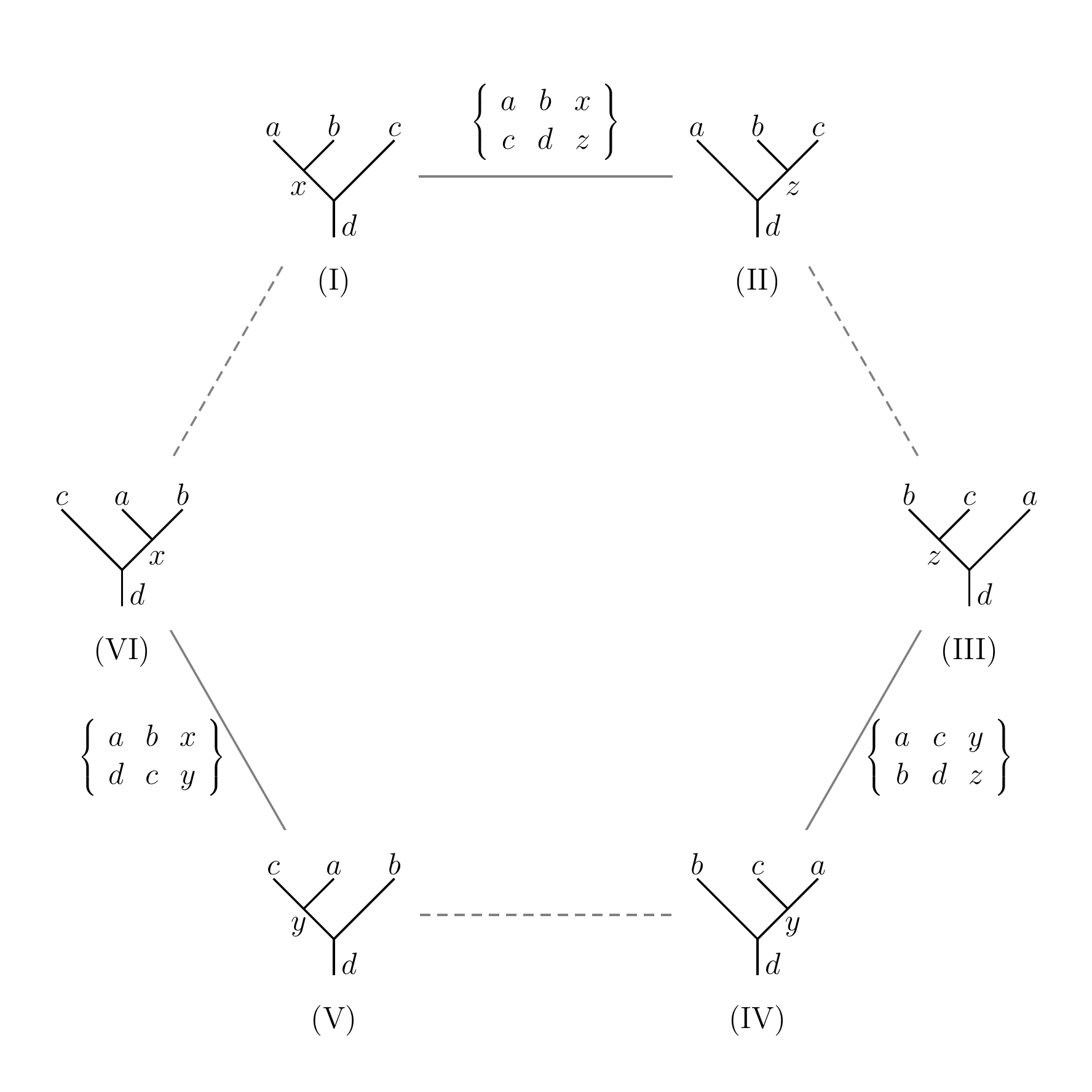}
\caption{\label{FIG.01} The hexagonal illustration of the Racah
sum rule, Eqs. \ref{eq02} and \ref{eq03}, showing relationship
among the three involved $6j$ symbols, indicating for each the
corresponding changes in coupling schemes. Transformations shown
as dotted lines involve simple phases only \cite{ac.01}.}
\end{figure}

We have that $x$, $y$, $z$ take $2a+1$ values, and
\begin{itemize}
    \item[$\bullet$] $x$ is $j_{ab}=j_{cd}$, range $[b-a,b+a]$
    \item[$\bullet$] $y$ is $j_{ac}=j_{bd}$, range $[c-a,c+a]$
    \item[$\bullet$] $z$ is $j_{ad}=j_{bc}$, range $[d-a,d+a]$
\end{itemize}
Specifically, $x$ and $z$ are the previous $j_{12}$ and $j_{23}$
\cite{lncs1.2013,lncs2.2013}, or $\ell$ and $\tilde{\ell}$ in the
previous paper on the volume operator \cite{ac.01}, and $y$ is the
\emph{seventh} Racah hidden momentum according to Labarthe
\cite{Lab2000}.

Based again on \emph{e.g.}  references  \cite{ac.01} and \cite{varsh} we list
here the three defining properties of the $6j$ symbols, according
to \cite{fanoracah1959}:
\begin{enumerate}
 \item \emph{Orthonormality}:
  \begin{equation}\label{eq01}
   \Sigma_x(2x+1)\iseij{a}{b}{x}{c}{d}{y}\iseij{c}{d}{x}{a}{b}{y'} = \frac{\delta_{yy'}}{2y'+1}\{a~ d~ y\}\{b~ c~ y \}
  \end{equation}
  where $\{...\}$ means that the three entries obey the triangular relationship.
 \item \emph{Additivity}: (Racah sum rule)
  \begin{equation}\label{eq02}
   \Sigma_x(-1)^{z+y+x}(2x+1)\iseij{a}{b}{x}{c}{d}{y}\iseij{c}{d}{x}{b}{a}{y} = \iseij{c}{a}{y}{d}{b}{z}
  \end{equation}
 \item \emph{Associativity}: The third (and last defining) property is the Biedenharn-Elliot identity. It provides
       the associative relationship schematically visualized as the pentagons in \cite{ac.01}.
       We remark here that from a projective viewpoint, it makes the underlying geometry Desarguesian \cite{fanoracah1959,robinson1970group,judd1983angularmomentum}.
\end{enumerate}

Using (\ref{eq01}), Eq. (\ref{eq02}) becomes (\cite{varsh}, eq.
(23) p. 467)
\begin{eqnarray}\label{eq03}
 \Sigma_{xy}(-1)^{x+y+z}(2x+1)(2y+1)\iseij{a}{b}{x}{c}{d}{z}\iseij{a}{c}{y}{d}{b}{x}\iseij{a}{d}{z'}{b}{c}{y}\nonumber\\
 = \frac{\delta_{zz'}}{2z+1}\{a~d~z\}\{b~c~z\}
\end{eqnarray}

The hexagonal illustration of the formulas, which involve seven
angular momenta, is in Figure \ref{FIG.01} (see also sec. 3.1 of
\cite{ac.01}).

The volume operator \cite{aquilanti2013volume,aquilanti2014}
\begin{eqnarray}
\hat{K}&=&\mathbf{J}_{a}\cdot\left(\mathbf{J}_{b}\times\mathbf{J}_{c}\right)
=
\mathbf{J}_{b}\cdot\left(\mathbf{J}_{c}\times\mathbf{J}_{d}\right)=\mathbf{J}_{a}\cdot\left(\mathbf{J}_{d}\times\mathbf{J}_{c}\right)=\mathbf{J}_{a}\cdot\left(\mathbf{J}_{b}\times\mathbf{J}_{d}\right)
\end{eqnarray}
acts
on $a$, $b$, $c$, $d$ independently of the order: we can expand
its eigenfunctions in one of the three alternative $|x \rangle$,
$|y \rangle$, and $|z \rangle$ bases \cite{cha1064}. We have of
course the same spectrum for any of the three choices, and the
wavefunctions are connected by orthogonal transformations
involving the three $6j$ symbols in Eqs. \ref{eq02} and
\ref{eq03}, and in Fig. \ref{FIG.01}.

\section{Results, exhibited as sequences of images}\label{s3}

\subsection{Caustics and Ridges}\label{s3.1}

\begin{figure}
  \center
  \subfigure[][$\iseij{30}{45}{x}{60}{55}{y}$]{
    \includegraphics[width=.45\textwidth, clip, trim= .5cm .5cm 1.7cm 1.7cm, page=1]{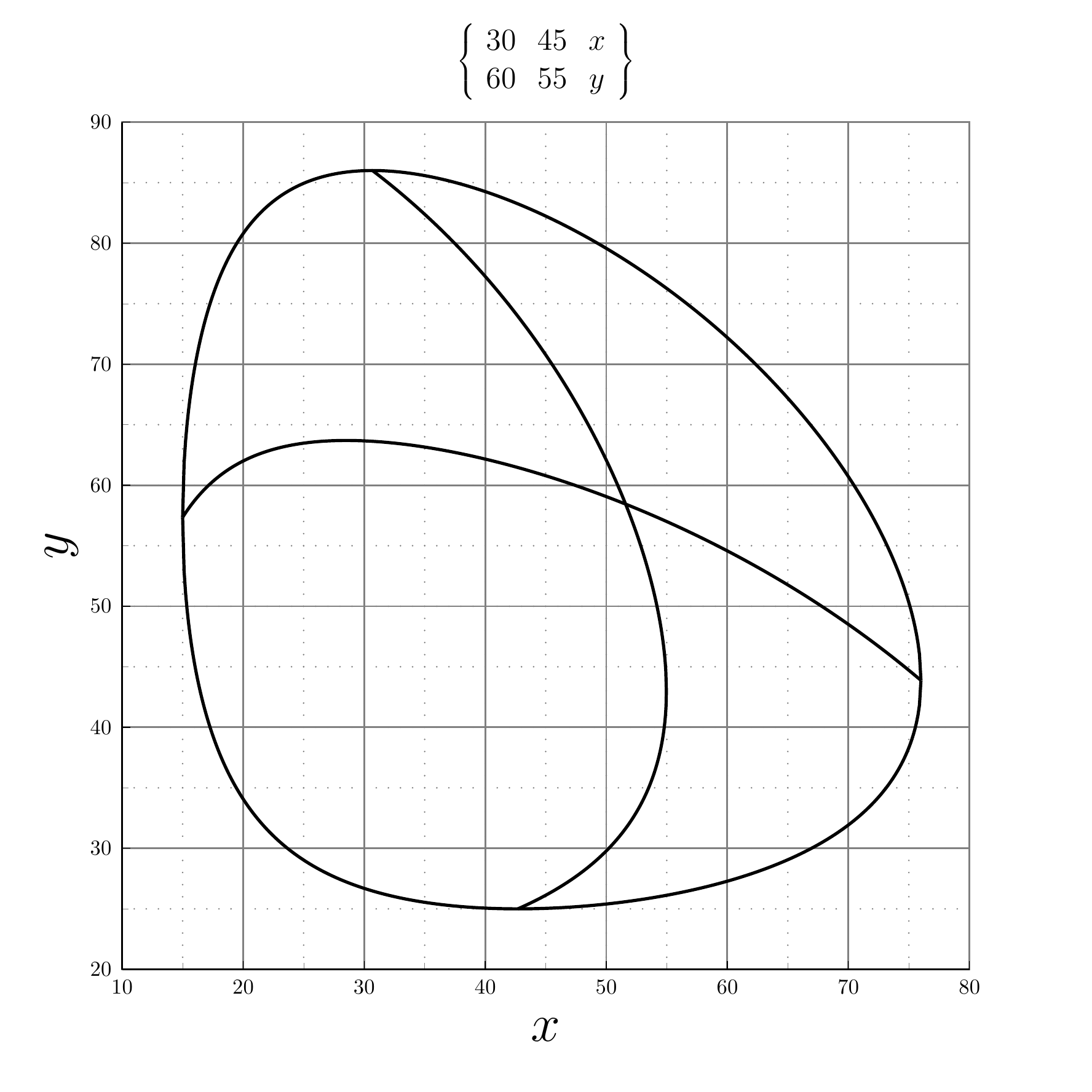}
  }
  \subfigure[][$\iseij{30}{55}{y}{45}{60}{z}$]{
    \includegraphics[width=.45\textwidth, clip, trim= .5cm .5cm 1.7cm 1.7cm, page=2]{FIGxyzA.pdf}
  }\\
  \subfigure[][$\iseij{30}{45}{x}{55}{60}{z}$]{
    \includegraphics[width=.45\textwidth, clip, trim= .5cm .5cm 1.7cm 1.7cm, page=3]{FIGxyzA.pdf}
  }
  \subfigure[][$xyz$ view]{\label{FIG.02d}
    \includegraphics[width=.45\textwidth, clip, trim= 1.5cm 1.5cm 1.5cm 1.5cm, page=4]{FIGxyzA.pdf}
  }
\caption{\label{FIG.02} Caustics and ridges of the indicated $6j$
symbols related by Racah sum rule Eqs. \ref{eq02} and \ref{eq03},
and Fig. \ref{FIG.01}. Fig. \ref{FIG.02d} gives the $xyz$ view.
General case, for which entries are chosen correspondingly to Fig.
1a in Ref. \cite{bitencourt2012exact}. Compare also with Fig. 1 in Ref.
\cite{lncs2.2013} where images are given of the $6j$ symbols
involved in the present panel (a). See the operator volume
potential functions $U^+$ and $U^-$ for this case in the Fig.
\ref{FIG.07} below.}
\end{figure}

\begin{figure}
  \center
  \subfigure[][\footnotesize$\iseij{100}{110}{x}{140}{130}{y}$]{
    \includegraphics[width=.45\textwidth, clip, trim= .5cm .5cm 1.7cm 1.7cm, page=1]{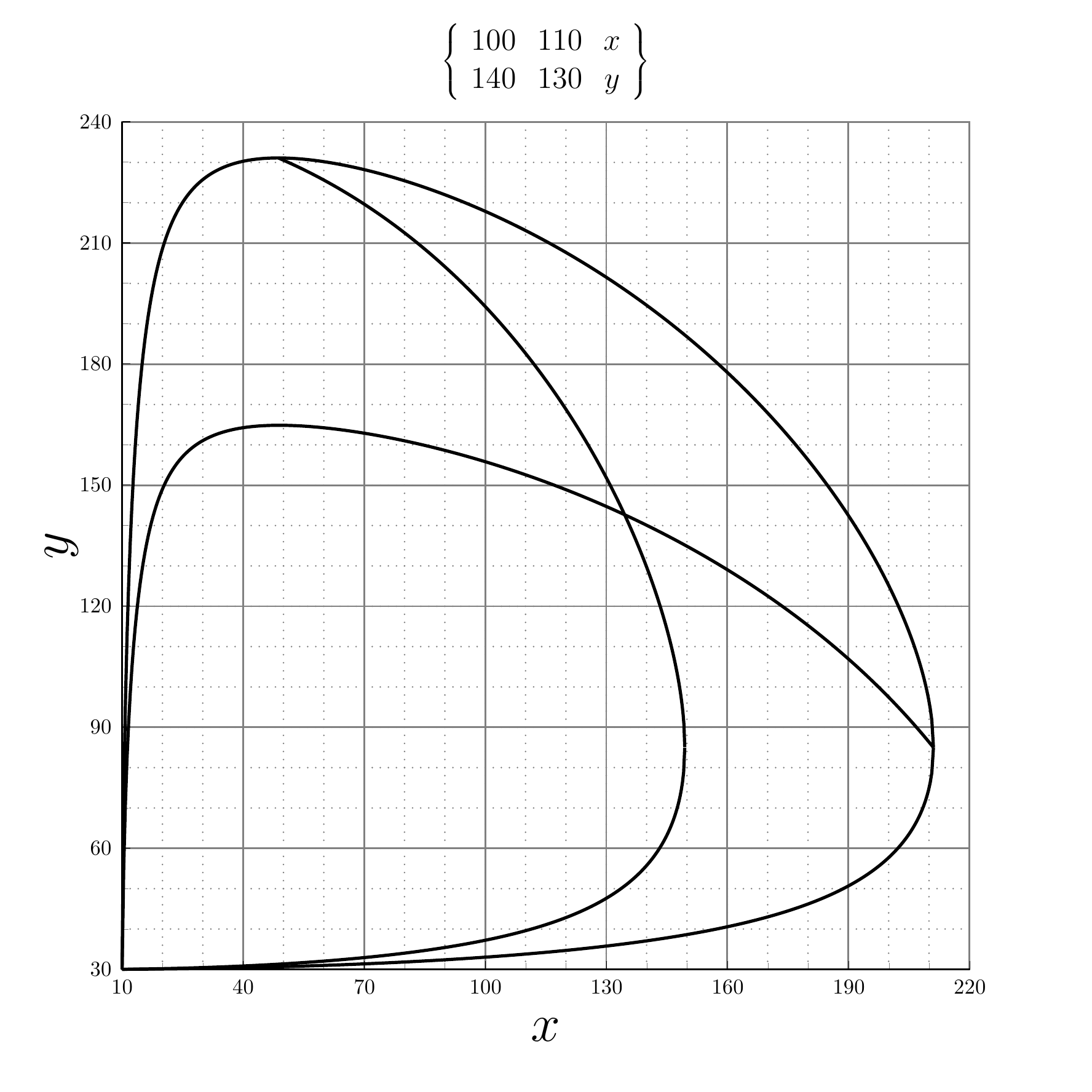}
  }
  \subfigure[][\footnotesize$\iseij{100}{130}{y}{110}{140}{z}$]{
    \includegraphics[width=.45\textwidth, clip, trim= .5cm .5cm 1.7cm 1.7cm, page=2]{FIGxyzB.pdf}
  }\\
  \subfigure[][\footnotesize$\iseij{100}{110}{x}{130}{140}{z}$]{
    \includegraphics[width=.45\textwidth, clip, trim= .5cm .5cm 1.7cm 1.7cm, page=3]{FIGxyzB.pdf}
  }
  \subfigure[][\footnotesize$xyz$ view]{
    \includegraphics[width=.45\textwidth, clip, trim= 1.5cm 1.5cm 1.5cm 1.5cm, page=4]{FIGxyzB.pdf}
  }
\caption{\label{FIG.03} As in Fig. \ref{FIG.02}, for a Regge
symmetric case. Entries correspond to Figs. 1b and 1c in Ref.
\cite{bitencourt2012exact}. Compare also with Figs. 5a and 5b in
Ref. \cite{lncs2.2013}, where images are given of the $6j$ symbols
involved in the present panel (c). Note coalescences as
characteristic of Regge symmetry, originating caustics cusps in the  different corners of
the screen for the various cases. See the operator volume
potential functions $U^+$ and $U^-$ for this case in the Fig.
\ref{FIG.08} below.}
\end{figure}

\begin{figure}
  \center
  \subfigure[][\footnotesize$\iseij{100}{100}{x}{200}{200}{y}$]{
    \includegraphics[width=.45\textwidth, clip, trim= .5cm .5cm 1.7cm 1.7cm, page=1]{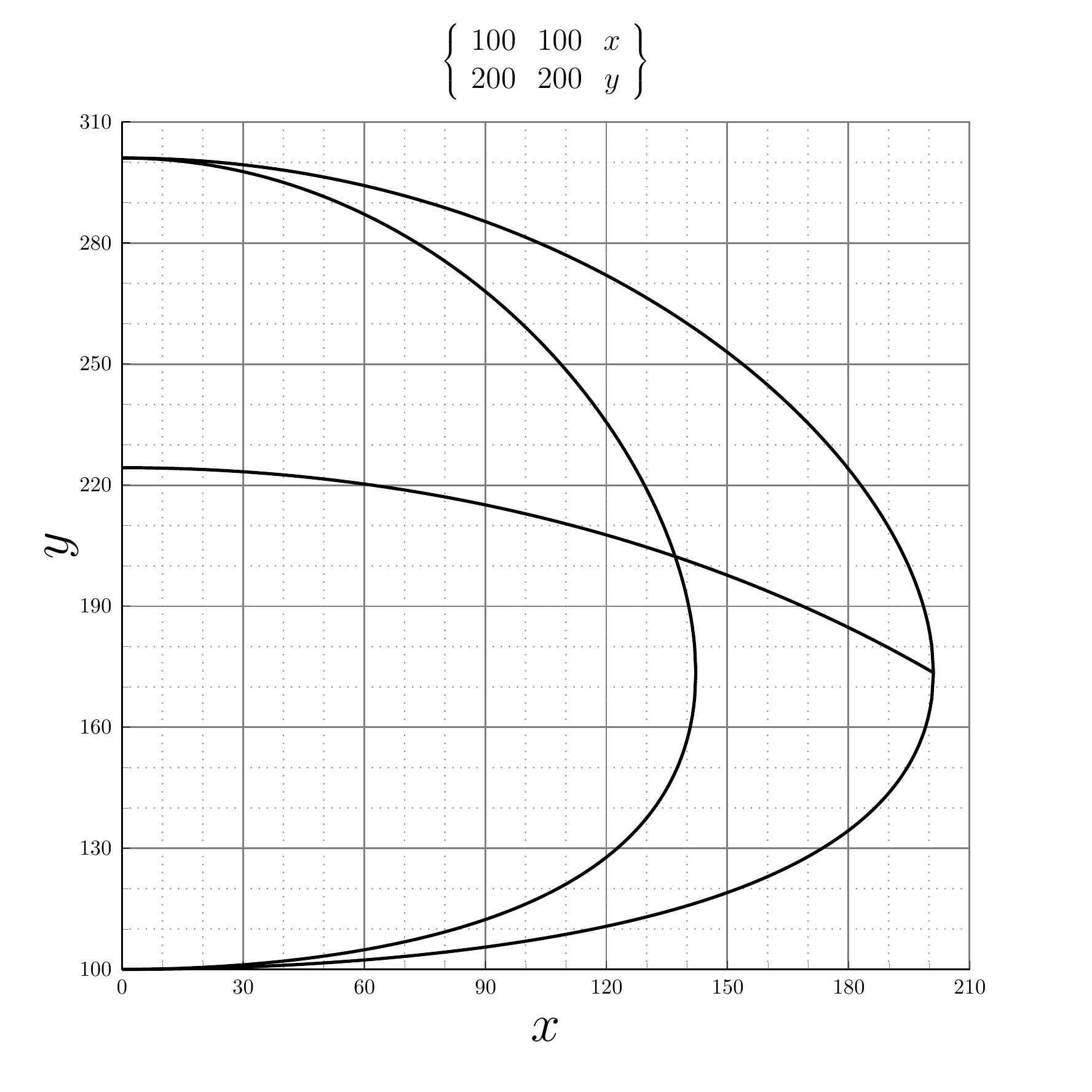}
  }
  \subfigure[][\footnotesize$\iseij{100}{200}{y}{100}{200}{z}$]{
    \includegraphics[width=.45\textwidth, clip, trim= .5cm .5cm 1.7cm 1.7cm, page=2]{FIGxyzC.pdf}
  }\\
  \subfigure[][\footnotesize$\iseij{100}{100}{x}{200}{200}{z}$]{
    \includegraphics[width=.45\textwidth, clip, trim= .5cm .5cm 1.7cm 1.7cm, page=3]{FIGxyzC.pdf}
  }
  \subfigure[][\footnotesize$xyz$ view]{
    \includegraphics[width=.45\textwidth, clip, trim= 1.5cm 1.5cm 1.5cm 1.5cm, page=4]{FIGxyzC.pdf}
  }
\caption{\label{FIG.04} As in Fig. \ref{FIG.02}, for a symmetric
degenerate case. See the operator volume potential functions $U^+$
and $U^-$ for this case in the Fig. \ref{FIG.09} below.}
\end{figure}

\begin{figure}
  \center
  \subfigure[][\footnotesize$\iseij{100}{200}{x}{200}{200}{y}$]{
    \includegraphics[width=.45\textwidth, clip, trim= .5cm .5cm 1.7cm 1.7cm, page=1]{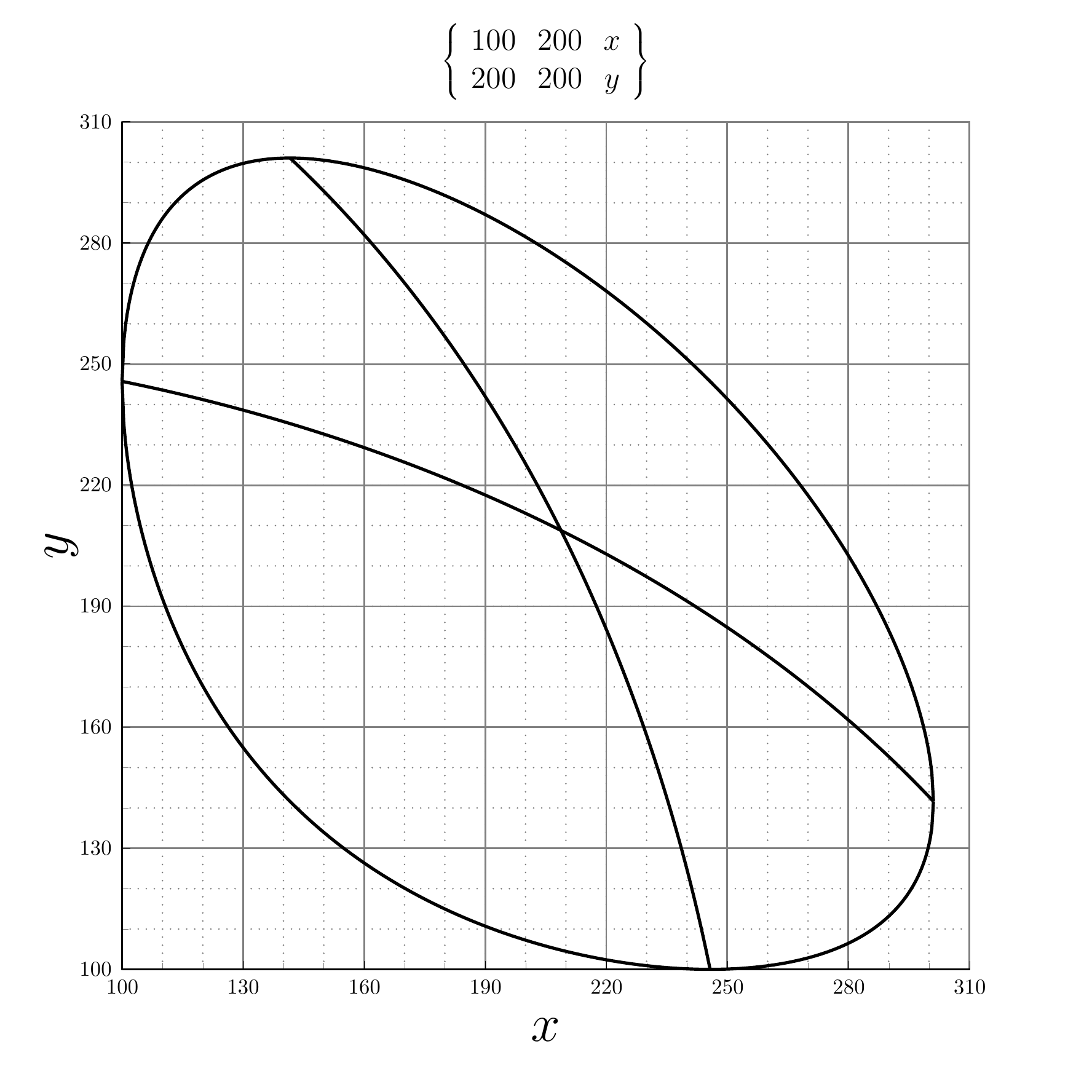}
  }
  \subfigure[][\footnotesize$\iseij{100}{200}{y}{200}{200}{z}$]{
    \includegraphics[width=.45\textwidth, clip, trim= .5cm .5cm 1.7cm 1.7cm, page=2]{FIGxyzD.pdf}
  }\\
  \subfigure[][\footnotesize$\iseij{100}{200}{x}{200}{200}{z}$]{
    \includegraphics[width=.45\textwidth, clip, trim= .5cm .5cm 1.7cm 1.7cm, page=3]{FIGxyzD.pdf}
  }
  \subfigure[][\footnotesize$xyz$ view]{
    \includegraphics[width=.45\textwidth, clip, trim= 1.5cm 1.5cm 1.5cm 1.5cm, page=4]{FIGxyzD.pdf}
  }
  \caption{\label{FIG.05} As in Fig. \ref{FIG.02}, for a further symmetric case.}
\end{figure}

\begin{figure}
  \center
  \subfigure[][\footnotesize$\iseij{100}{100}{x}{100}{100}{y}$]{
    \includegraphics[width=.45\textwidth, clip, trim= .5cm .5cm 1.7cm 1.7cm, page=1]{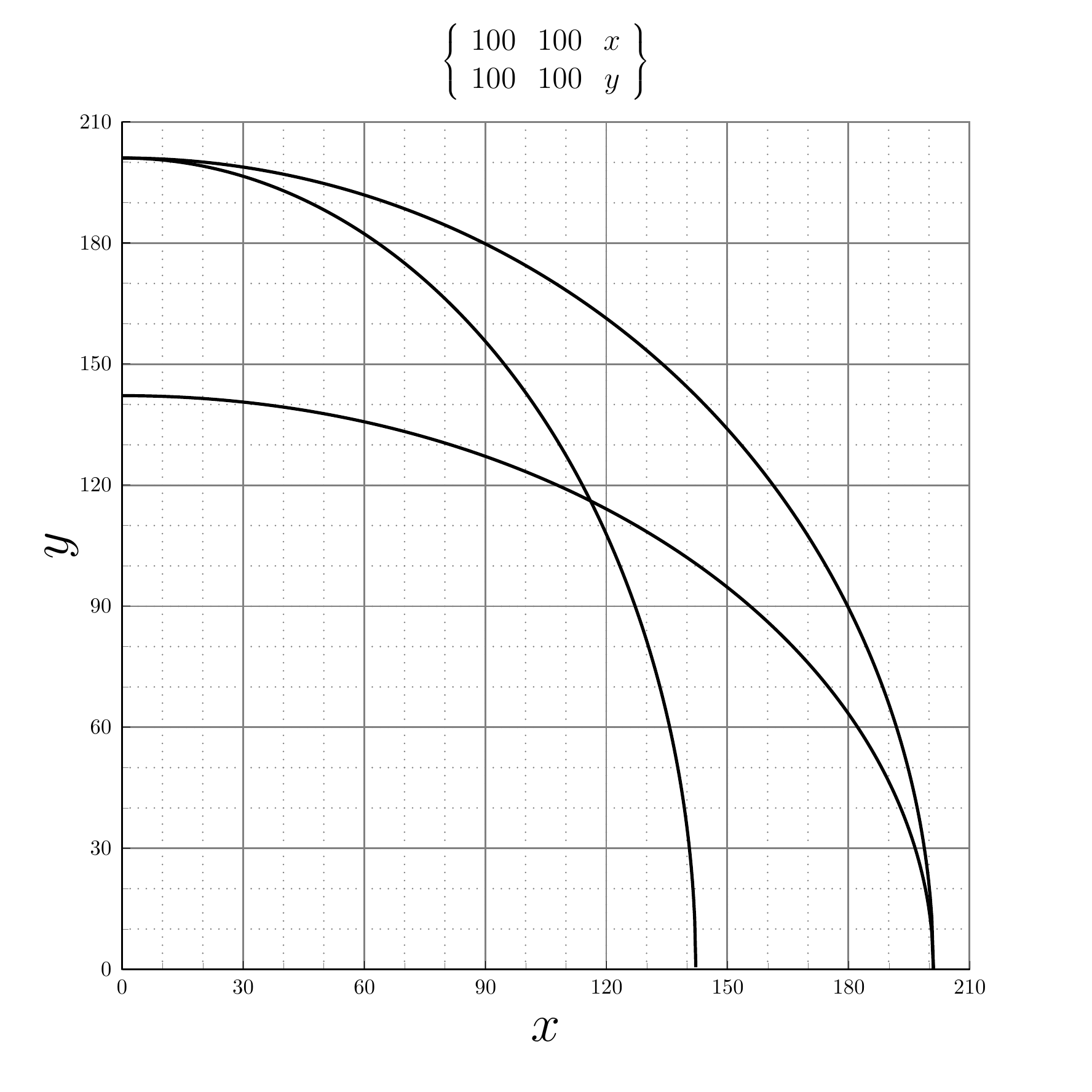}
  }
  \subfigure[][\footnotesize$\iseij{100}{100}{y}{100}{100}{z}$]{
    \includegraphics[width=.45\textwidth, clip, trim= .5cm .5cm 1.7cm 1.7cm, page=2]{FIGxyzE.pdf}
  }\\
  \subfigure[][\footnotesize$\iseij{100}{100}{x}{100}{100}{z}$]{
    \includegraphics[width=.45\textwidth, clip, trim= .5cm .5cm 1.7cm 1.7cm, page=3]{FIGxyzE.pdf}
  }
  \subfigure[][\footnotesize$xyz$ view]{
   \includegraphics[width=.45\textwidth, clip, trim= 1.5cm 1.5cm 1.5cm 1.5cm, page=4]{FIGxyzE.pdf}
  }
\caption{\label{FIG.06} As in Fig. \ref{FIG.02}, for a fully
symmetric case. This is to be compared with Fig. 6 in Ref.
\cite{bitencourt2012exact} and with the images in Fig. 9 in
\cite{lncs2.2013}. Corresponding images for the eigenfunctions of
the volume operator for the fully symmetric case are in Fig.
\ref{FIG.10} of the present paper.}
\end{figure}

For five sets of the $a, b, c, d$ parameter, chosen in line with
previous papers in this series
\cite{bitencourt2012exact,lncs1.2013,lncs2.2013}, Figs.
\ref{FIG.02}-\ref{FIG.06} illustrate caustics and ridges by means
of three panels for each of the three corresponding $6j$ symbols.

For each case an additional panel shows an orthogonal three
dimensional projection. Plots in panels are referred to as $xz$,
$yz$, and $xy$ views, which are sections of surfaces of constant
volume, that we call Piero's ``eggs" (see remark (iii) in Sec.
\ref{s4}). Specifically, the $xz$ $6j$, the $xy$ $6j$, and the
$yz$ $6j$ are
\begin{equation}
\iseij{a}{b}{x}{c}{d}{z}~,~~~~ \iseij{a}{b}{x}{d}{c}{y}~,~~~~
\iseij{a}{c}{y}{b}{d}{z}
\end{equation}

\begin{figure}[h]
  \center
  \includegraphics[width=0.8\textwidth]{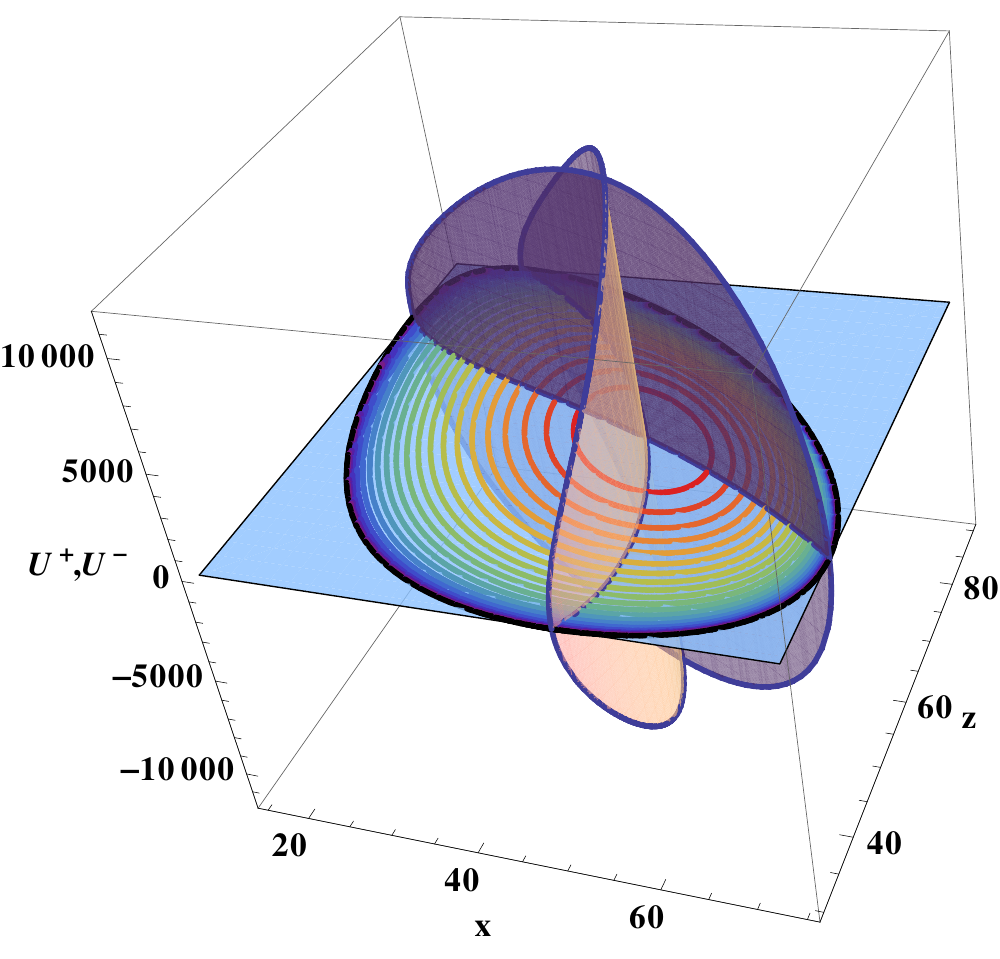}
\caption{\label{FIG.07} Operator volume potential functions
\cite{aquilanti2013volume} $U^+$ for positive values and $U^-$ for
negative values (partially hidden) for $a=30$, $b=45$, $c=55$,
$d=60$. This is the same as considered in Fig. \ref{FIG.02}, the
horizontal zero plane here corresponding to panel (c) there.}
\end{figure}

The various sets in Figs. \ref{FIG.02}-\ref{FIG.07} illustrate a
generic case and permutational or Regge symmetrical cases, with
parameters taken from previous papers \cite{lncs1.2013,lncs2.2013} to permit comparisons.
\newpage

\subsection{Potential $U^+$ and $U^-$ and eigenfunctions for the volume operator}\label{s3.2}

Potential functions for the volume operator, $U^+$ and $U^-$,
correspond to sections along ridges of the ``egg"
\cite{aquilanti2013volume}. They are shown for some of the cases
considered in Sec. \ref{s3.1} in Fig. 7-9.

Since $U^+ = - U^-$,
these figures are symmetric with respect to the planes (the ``screens'') of the
previous figures. Finally a plot of the volume operator
wavefunctions is reported in Fig. \ref{FIG.10} for a symmetrical case.

\begin{figure}
  \center
  \includegraphics[width=0.8\textwidth]{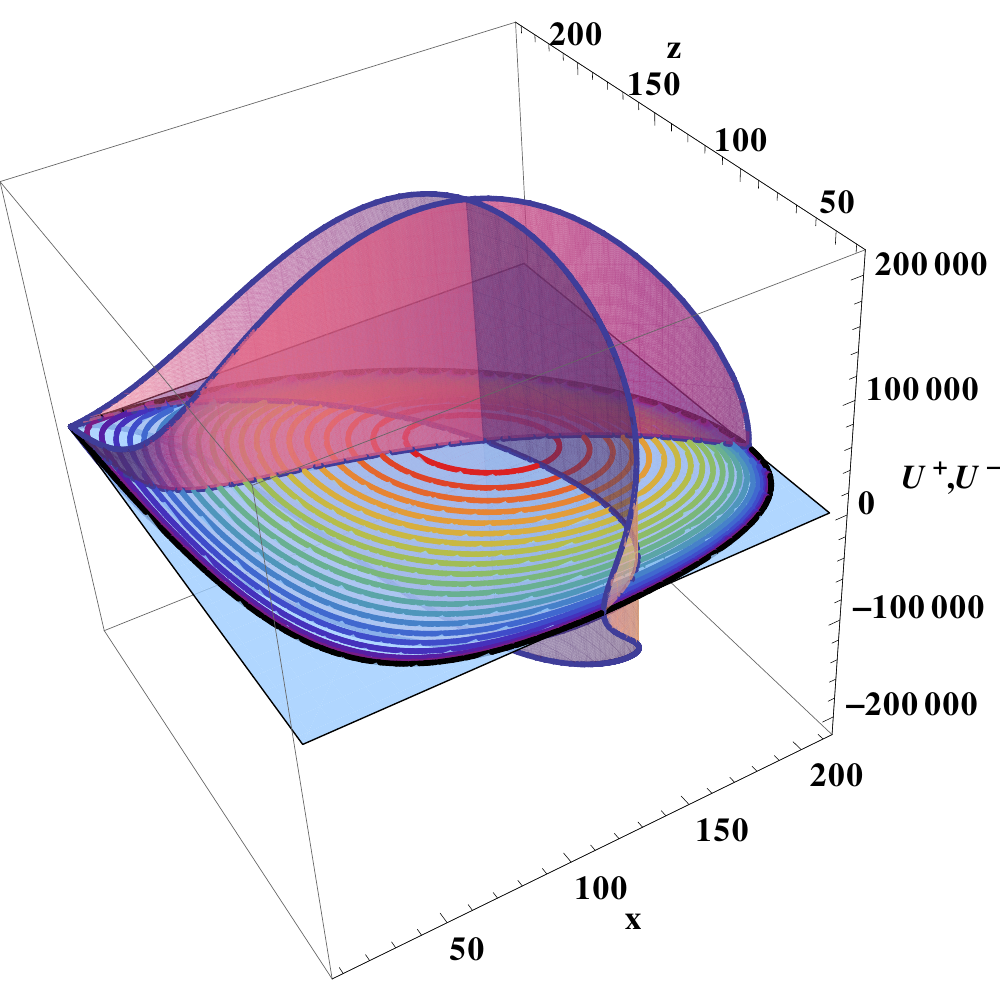}
\caption{\label{FIG.08} Operator volume potential functions
\cite{aquilanti2013volume} $U^+$ for positive values and $U^-$
(partially hidden) for negative values, for $a=100$, $b=110$,
$c=130$, $d=140$. This is the case considered in Fig.
\ref{FIG.03}, the horizontal zero plane here corresponding to
panel (c) there.}
\end{figure}
\begin{figure}
  \center
  \includegraphics[width=0.8\textwidth]{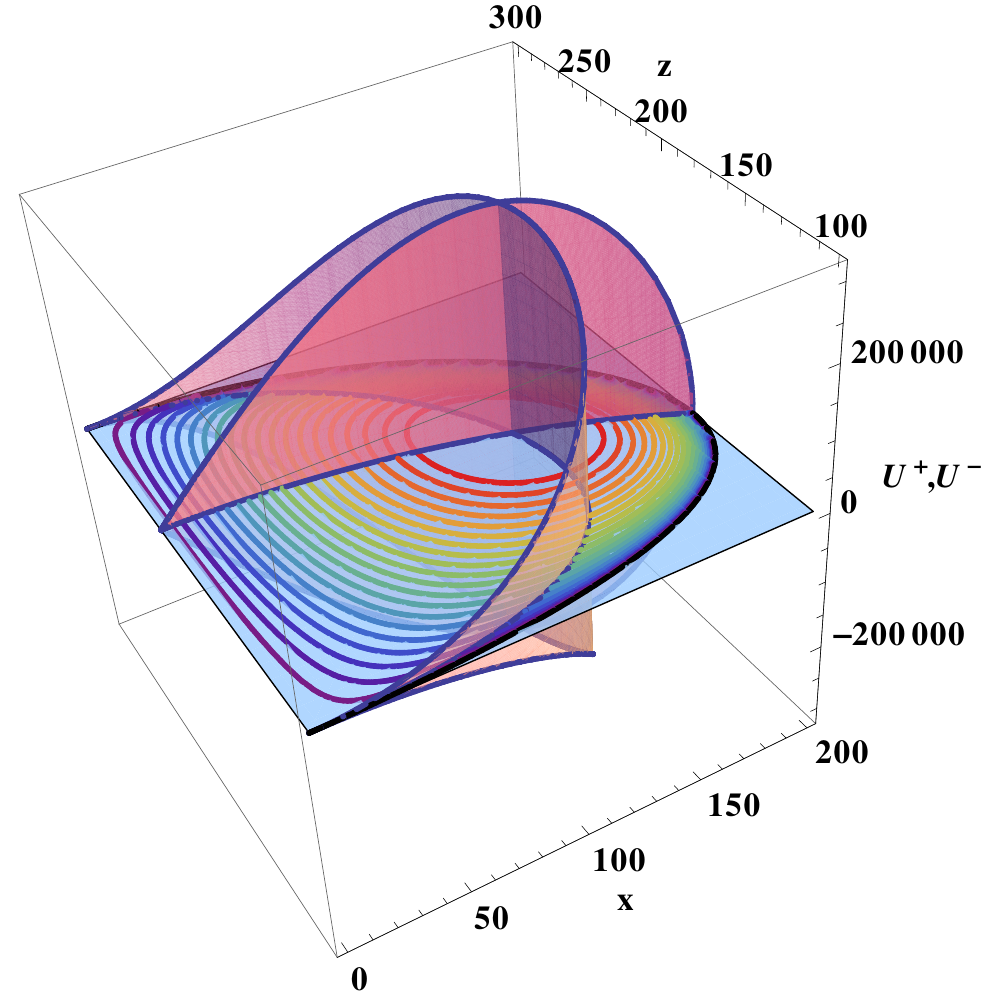}
  \caption{\label{FIG.09} As for previous figure, for $a=b=100$, $c=d=200$. This is the same
case considered in Fig. \ref{FIG.04}, the horizontal zero plane
here corresponding to panel (a) there.}
\end{figure}
\begin{figure}[ht]
  \center
  \includegraphics[width=0.8\textwidth]{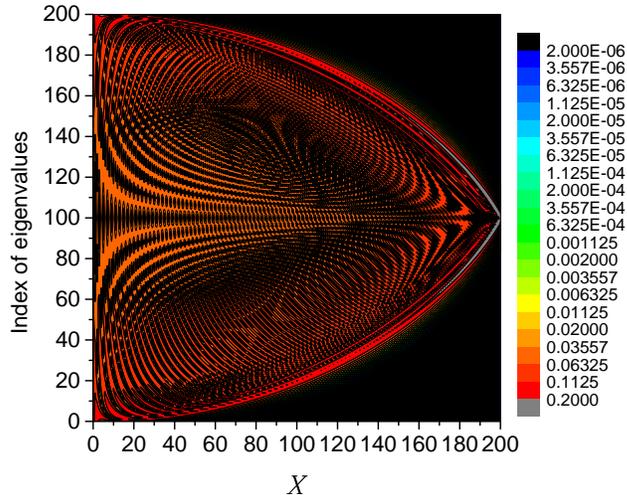}
\caption{\label{FIG.10} The eigenfunctions of the volume operator,
as defined in Ref. \cite{aquilanti2013volume}, for the fully
symmetric case, for which  four spins $j_1 \ldots j_4$ are equal
and also large. According to the nomenclature in this paper,
$a=b=c=d= 100$. In the abscissa we have what we called $l$ in Ref.
\cite{aquilanti2013volume} and $x$ in this paper. The ordinate
scale is the index of eigenvalues (0 for the lowest, etc.), that
are the same in number (201) as for $x$. They are symmetric with
respect to the 101$st$ eigenvalue, the fast oscillations having
been suppressed since here absolute values are reported.
Comparison can be made with plots for these values of the
parameters with the caustics and ridges of $6j$s in Fig.6 in this
paper and especially with the images in  Fig.9 in Ref.
\cite{lncs2.2013}.}
\end{figure}

\newpage
\section{Remarks}\label{s4}

\begin{itemize}
    \item [(i)] The four angular momenta $a, b, c, d$, and those arising
in their intermediate couplings, $x, y, z$, form what we define a
7-spin network, emphasizing the finite projective geometry
interpretation introduced by Robinson \cite{robinson1970group},
see also Refs. \cite{bilo9,Lab2000}. The fully symmetric nature of the
network involves mapping on the  Fano plane, introduced by G. Fano in 1892 as the smallest nontrivial finite projective geometry.
    \item [(ii)] From this viewpoint, the associative property in
Sec. \ref{s2} permits to move from the plane to space, and to
prove that the latter is Arguesian. Work is in progress to extend
the theory to higher spin-networks, in connection to the
morphogenetic approach of Ref. \cite{rva2009}.
    \item [(iii)] One of the greatest mathematical achievements of
    the Renaissance master Piero della Francesca ($\sim$ 1415 - 1492) was
    the discovery of the formula for the volume of a generic
    tetrahedron in terms of its edge lengths, used here. In one of the most
    beautiful of his famous masterpieces (1472), he painted an
    enigmatic egg at the focus of a dome designed according the
    newly-born theory of perspective, of which he was one of the
    founding fathers. In our $xy$, $xz$, $zy$ plots of the volume
    surfaces corresponding to his formula one has ovaloid, egg -
    like shapes: caustics, ridges and the volume operator potentials $U^+$
    and $U^-$ are sections of those surfaces. There is no documentation that he was aware of such a connection,  which anyway  motivated us to call the ovaloid as Piero's     ``egg".
\end{itemize}

\bibliographystyle{splncs}

\begin{thebibliography}{10}

\bibitem{aquilanti2013volume}
Aquilanti, V., Marinelli, D., Marzuoli, A.:
\newblock Hamiltonian dynamics of a quantum of space: hidden symmetries and
  spectrum of the volume operator, and discrete orthogonal polynomials.
\newblock J. Phys. A: Math. Theor. \textbf{46} (2013)  175303 ~
  arXiv:1301.1949v1 [math-ph].

\bibitem{aquilanti2014}
Aquilanti, V., Marinelli, D., Marzuoli, A.:
\newblock Symmetric coupling of angular momenta, quadratic algebras and
  discrete polynomials.
\newblock J. Phys: Conf. Series \textbf{482} (2014)  012001 ~ arXiv:1401.3591v1
  [quant-ph].

\bibitem{cha1064}
Chakrabarti, A.:
\newblock On the coupling of 3 angular momenta.
\newblock Ann. Inst. H. Poincar\'e Sect. A \textbf{1} (1964)  301--327

\bibitem{jeanmarclevyleblond1965symmetrical}
L\'{e}vy-Leblond, J.M., L\'{e}vy-Nahas, M.:
\newblock {Symmetrical Coupling of Three Angular Momenta}.
\newblock J. Math. Phys. \textbf{6}(9) (1965)  1372--1380

\bibitem{car2002}
Carbone, G., Carfora, M., Marzuoli, A.:
\newblock Quantum states of elementary three-geometry.
\newblock Classical Quant. Grav. \textbf{19} (2002)  3761 ~
  arXiv:gr-qc/0112043.

\bibitem{granovskii19921}
Granovskii, Y., Lutzenko, I., Zhedanov, A.:
\newblock Mutual integrability, quadratic algebras, and dynamical symmetry.
\newblock Annals of Physics \textbf{217}(1) (1992)  1 -- 20

\bibitem{gen2014}
Genest, V., Vinet, L., Zhedanov, A.:
\newblock The equitable racah algebra from three su(1,1) algebras.
\newblock J. Phys. A: Math. Theor. \textbf{47} (2014)  025203 ~ arXiv:1309.3540
  [math-ph].

\bibitem{ter2001}
Terwilliger, P.:
\newblock Two linear transformations each tridiagonal with respect to an
  eigenbasis of the other.
\newblock Linear Algebra Appl. \textbf{30} (2001)  149--203 ~
  arXiv:math/0406555 [math.RA].

\bibitem{ponzregge}
Ponzano, G., Regge, T.:
\newblock Semiclassical limit of {R}acah coefficients.
\newblock Spectroscopic and Group Theoretical Methods in Physics (1968) F.
  Bloch et al (Eds.), North--Holland, Amsterdam, pp. 1-58.

\bibitem{neville2006volume1}
Neville, D.E.:
\newblock Volume operator for spin networks with planar or cylindrical
  symmetry.
\newblock Phys. Rev. D \textbf{73}(12) (June 2006)  124004 ~
  arXiv:gr-qc/0511005.

\bibitem{neville2006volume2}
Neville, D.E.:
\newblock Volume operator for singly polarized gravity waves with planar or
  cylindrical symmetry.
\newblock Phys. Rev. D \textbf{73} (June 2006)  124005 ~ arXiv:gr-qc/0511006.

\bibitem{neville1971technique}
Neville, D.E.:
\newblock A technique for solving recurrence relations approximately and its
  application to the {3-J} and {6-J} symbols.
\newblock J. Math. Phys. \textbf{12}(12) (1971)  2438--2453

\bibitem{nikiforovsuslovuvarov199110}
Nikiforov, A.F., Suslov, S.K., Uvarov, V.B.:
\newblock Classical Orthogonal Polynomials of a Discrete Variable (Scientific
  Computation).
\newblock Springer-Verlag (10 1991)

\bibitem{schgora}
Schulten, K., Gordon, R.:
\newblock Exact recursive evaluation of 3j- and 6j-coefficients for quantum
  mechanical coupling of angular momenta.
\newblock J. Math. Phys. \textbf{16} (1975)  1961--1970

\bibitem{schgorb}
Schulten, K., Gordon, R.:
\newblock Semiclassical approximations to 3j- and 6j-coefficients for
  quantum-mechanical coupling of angular momenta.
\newblock J. Math. Phys. \textbf{16} (1975)  1971--1988

\bibitem{rbfaal.10}
Ragni, M., Bitencourt, A., da~S.~Ferreira, C., Aquilanti, V.,
Anderson, R.,
  Littlejohn, R.:
\newblock Exact computation and asymptotic approximation of $6j$ symbols.
  illustration of their semiclassical limits.
\newblock Int. J. Quantum Chem. (110) (2010)  731--742

\bibitem{littlejohn2009uniform}
Littlejohn, R.G., Yu, L.:
\newblock Uniform semiclassical approximation for the {W}igner 6j-symbol in
  terms of rotation matrices.
\newblock J. Phys. Chem. A \textbf{113} (2009)  14904--14922 ~ arXiv:0905.4240
  [math-ph].

\bibitem{aquilanti2007semiclassical}
Aquilanti, V., Haggard, H.M., Littlejohn, R.G., Yu, L.:
\newblock {Semiclassical analysis of {W}igner $3j$ -symbol}.
\newblock J. Phys. A \textbf{40}(21) (2007)  5637--5674 ~
  arXiv:quant-ph/0703104.

\bibitem{fanoracah1959}
Fano, U., Racah, G.:
\newblock {Irreducible tensorial sets}. 1st edn. Volume~4 of Pure and applied
  physics.
\newblock Academic Press (1959)

\bibitem{robinson1970group}
de~B.~Robinson, G.:
\newblock {Group Representations and Geometry}.
\newblock J. Math. Phys. \textbf{11}(12) (1970)  3428--3432

\bibitem{judd1983angularmomentum}
Judd, B.:
\newblock {Angular-momentum theory and projective geometry}.
\newblock Found. Phys. \textbf{13}(1) (January 1983)  51--59

\bibitem{bilo9}
Biedenharn, L.C., Louck, J.D.:
\newblock The {R}acah--{W}igner Algebra in Quantum Theory.
\newblock Number Rota, G--C. (Ed) in Encyclopedia of Mathematics and its
  Applications Vol 9. Addison--Wesley Publ. Co.: Reading MA (1981)

\bibitem{Lab1975}
Labarthe, J.J.:
\newblock Generating functions for the coupling-recoupling coefficients of
  su(2).
\newblock J. Phys. A \textbf{8}(10) (1975)  1543

\bibitem{Lab1998}
Labarthe, J.J.:
\newblock {The hidden angular momenta of Racah and $3n-j$ coefficients}.
\newblock J. Phys. A \textbf{31} (1998)  8689

\bibitem{Lab2000}
Labarthe, J.J.:
\newblock The hidden angular momenta for the coupling-recoupling coefficients
  of su(2).
\newblock J. Phys. A \textbf{33} (2000)  763

\bibitem{bitencourt2012exact}
Bitencourt, A.C., Marzuoli, A., Ragni, M., Anderson, R.W.,
Aquilanti, V.:
\newblock Exact and asymptotic computations of elementary spin networks:
  Classification of the quantum-classical boundaries.
\newblock In: Lecture Notes in Computer Science. Volume I-7333., Springer
  (2012)  723--737 ~ arXiv:1211.4993[math-ph].

\bibitem{lncs1.2013}
Anderson, R.W., Aquilanti, V., Bitencourt, A.C.P., Marinelli, D.,
Ragni, M.:
\newblock The screen representation of spin networks: 2d recurrence, eigenvalue
  equation for 6j symbols, geometric interpretation and hamiltonian dynamics.
\newblock Lecture Notes in Computer Science \textbf{7972} (2013)  46--59 ~
  arXiv:1404.4555 [quant-ph].

\bibitem{lncs2.2013}
Ragni, M., Littlejohn, R.G., Bitencourt, A.C.P., Aquilanti, V.,
Anderson, R.W.:
\newblock The screen representation of spin networks. images of $6j$ symbols
  and semiclassical features.
\newblock Lecture Notes in Computer Science \textbf{7972} (2013)  60--72 ~
  arXiv:1405.0837 [quant-ph].

\bibitem{ac.01}
Aquilanti, V., Coletti, C.:
\newblock $3nj$-symbols and harmonic superposition coefficients: an icosahedral
  abacus.
\newblock Chem. Phys. Letters \textbf{344} (2001)  601--611

\bibitem{varsh}
Varshalovich, D., Moskalev, A., Khersonskii, V.:
\newblock Quantum Theory Of Angular Momentum.
\newblock World Scientific, Singapore (10 1988)

\bibitem{rva2009}
Anderson, R.W., Aquilanti, V., Marzuoli, A.:
\newblock 3nj morphogenesis and semiclassical disentangling.
\newblock J. Phys. Chem. A \textbf{113} (2009)  15106 -- 15117 ~
  arXiv:1001.4386 [quant-ph].

\end{thebibliography}

\end{document}